\documentclass[aps,prb,notitlepage,12pt]{revtex4-1} 

\usepackage{amsmath}  
\usepackage{amsfonts} 
\usepackage{graphicx} 

\usepackage[colorlinks]{hyperref}
\hypersetup{citecolor=blue}
\hypersetup{linkcolor=blue}
\hypersetup{urlcolor=blue}

\begin{document}

\title{New measurements of a simple pendulum using acceleration sensors.}

\author{Julien Vandermarli\`ere}
\affiliation{Saint-Exupery Secondary School, 13 avenue Paul Alduy,
66100 Perpignan, France; julien.vandermarliere@ac-montpellier.fr}

\author{Mikhail Indenbom}
\affiliation{Department of Physics, University of West Brittany, 
6 avenue Le Gorgeu, 29200 Brest, France; mikhail.indenbom@univ-brest.fr}


\maketitle 

To measure oscillation of a simple pendulum was probably a first idea coming to mind after appearance of smartphones with small but powerful acceleration 
sensors~: Simply attach the telephone to a playground swing\cite{pendrill1} 
or hang it on two string as the pendulum bob\cite{vogt1} and record the data. 
But immediately the problem becomes not so trivial. To deeply  investigate on it\cite{Fernandes}
or to make the phase diagram of the movement\cite{Monteiro} require complex calculations 
that are far beyond the capability of the youngest students. In this article we propose another
way to study the pendulum by putting the sensors at the axis of rotation. This method gives 
immediately the figures very close to ones in textbooks. 

\section*{Theoretical background}

For a simple pendulum, where all mass is concentrated in its bob, the 
tangential acceleration $L\ddot{\theta}$ is caused by the component of the gravitational
force in this direction $-mg\sin\theta$ ($m$ is the pendulum mass, $L$ is its length and
$\theta$, $\dot{\theta}$, $\ddot{\theta}$ are the deviation angle and its first and second
derivatives by time, correspondingly). So, one would expect to measure this acceleration
using an accelerometer put on the pendulum bob. But the accelerometer  measures not the 
acceleration vector $\vec{a}$ but the vector $\vec{a}-\vec{g}$ ($\vec{g}$ is the gravitational 
acceleration of a free fall pointing vertically down). 
Thus the measured value in the direction of motion $x$ is zero\cite{pendrill1}~:
\begin{equation}
	A_x = L\ddot{\theta} + g\sin\theta = 0\ .
\end{equation}
Additional forces
of friction and a more complicated mass distribution (physical pendulum) give some, 
usually small, deviations of $A_x$ from zero.\cite{Fernandes} 

The indication of the accelerometer along the pendulum length is created by 
the centripetal acceleration and the axial component of gravitation~:
\begin{equation}
    A_y =  L\dot{\theta}^2 + g\cos\theta \ .
\end{equation}
Considering small oscillations $\theta = \theta_0\cos(\omega t)$ and taking into account
$\cos\theta \approx 1 - \frac{1}{2}\theta^2$ we get that $A_y$ oscillates near $g$
with \emph{double frequency} and an amplitude proportional to $\theta_0^2$. Thus
the signal should disappear very fast with reduction of $\theta_0$. All these facts
make measurements of $A_y$ not very appropriate for characterization of the pendulum motion.

If one wants to use the accelerometer for its original purpose in smartphones~: to determine
the orientation of it relatively to vertical $\vec{g}$, he should place the probe so that it
does not move and only rotates, namely, in our case, \emph{at the pendulum rotation axis}. 
In this position the accelerometer will indicate $a_x = g\sin\theta$ and 
$a_y = g\cos\theta$ ($y$ is always pointing up along the pendulum) giving direct information
on the rotation angle $\theta$. For small deviations of the pendulum from the vertical
$\theta \approx a_x/g$ simply.

\section*{The experiment}

\begin{figure}[!ht]
\centering
\includegraphics[width=3in]{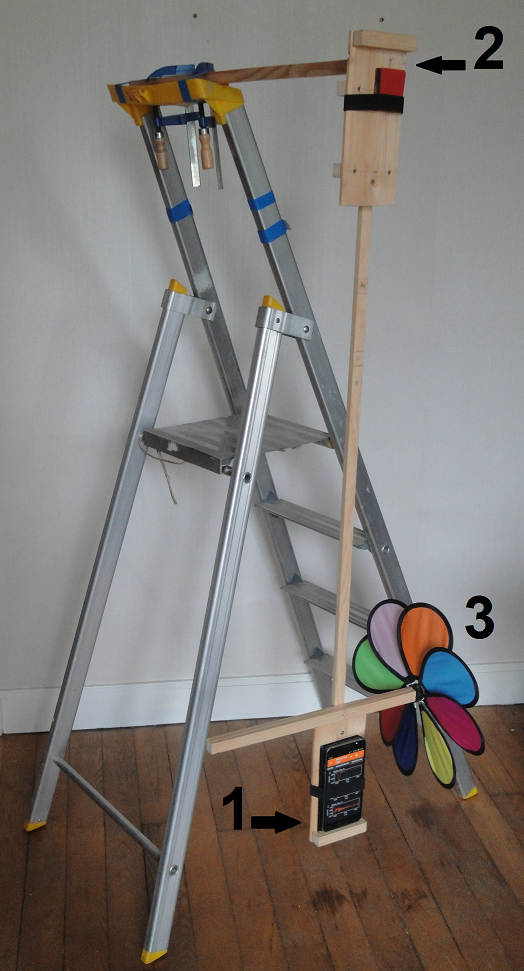}
\caption{Pendulum with two points of measurements. (1)~: A smartphone used as the
pendulum bob which measures ordinary effects. (2)~: A SensorTag measuring at the same time
the pendulum deviation $\theta$ and its rotation speed $\dot{\theta}$. A colorful helix (3)
was added to increase the dumping of oscillations.}
\label{pendulum}
\end{figure}

In order to illustrate the above statements we have created a pendulum where the 
smartphone can be placed in two positions: down as the pendulum bob,  and up so, that
the location of the accelerometer\cite{place,place3} coinsides with the rotation axis 
(Fig. \ref{pendulum}). 
 For being more clear we have used an additional SensorTag (CC2650STK of Texas Instruments)
in order to have measurements in two positions at the same time.

The parallel recordings of two sensors were realized using phyphox 
software.\cite{nature,phyphox}. Using this software and numerous examples given in
free access one can create his own complicated experiments with different sensors
at the same time and easily save obtained data for further treatment, 
by means of Python in our case.

\begin{figure}[!ht]
\centering
\includegraphics[width=6in]{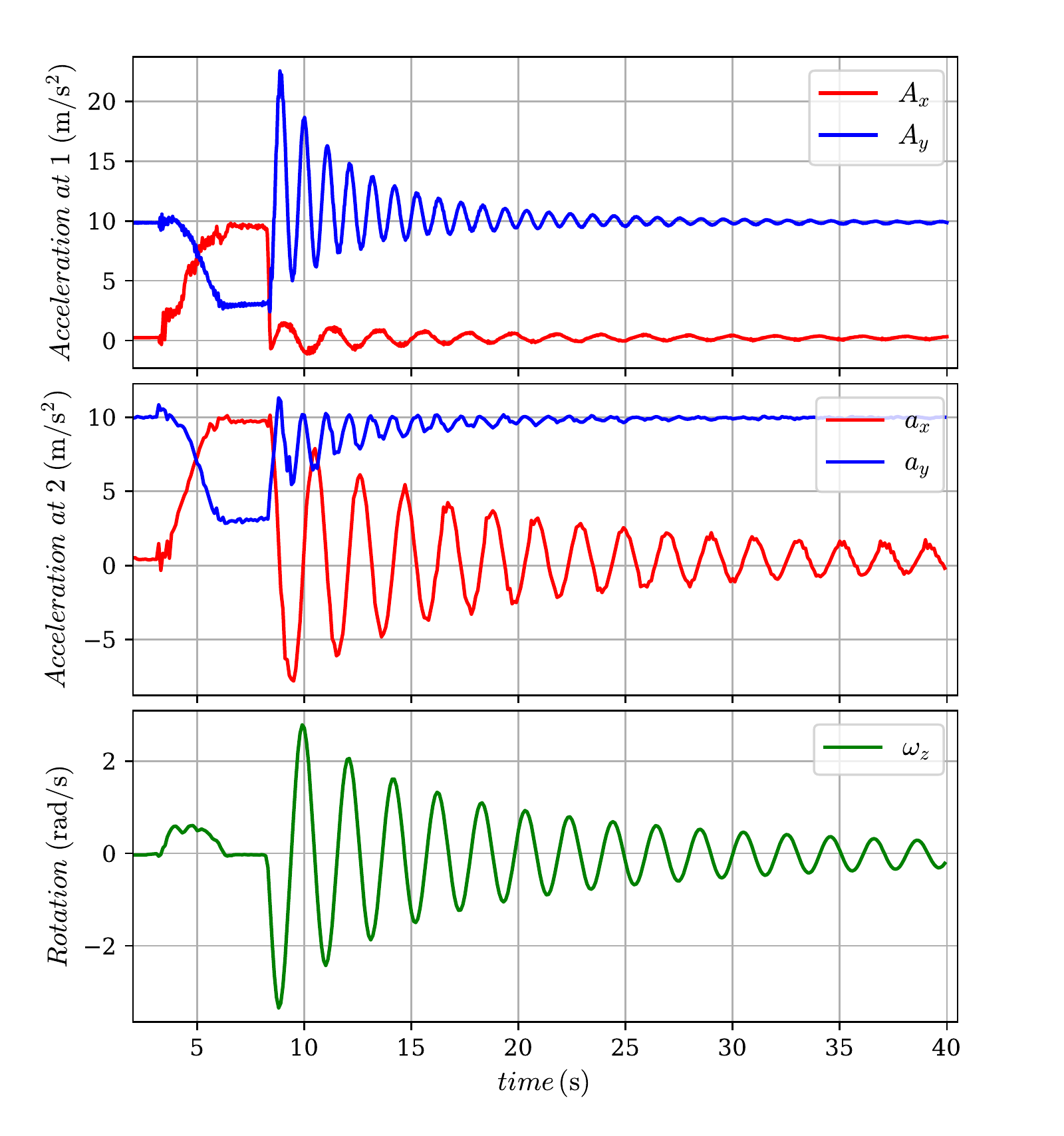}
\caption{Measurements by the acceleration sensors at different points. Upper panel~:
The sensor (here of a smartphone) is in the usual position at the pendulum bob.
$A_x$ measured in the motion direction 
drops to zero immediately after releasing the pendulum ($t \approx 8\rm\, s$).
Central panel~: Parallel measurements at the pendulum rotation axis (here by a SensorTag). 
$a_x$ follows well the oscillations of the deviation angle $\theta$.
Lower panel: Pendulum rotation speed $\dot{\theta}$
(gyroscope component $\omega_z$ perpendicular to the 
oscillation plane) measured at the same time.}
\label{3graphs}
\end{figure}

The results presented here are obtained with the smartphone in the down (usual) position and 
the SensorTag in the upper position at the rotation axis as can be seen in Fig.~\ref{pendulum}. 
The smartphone accelerometer demonstrates ordinary curves (upper panel in Fig.~\ref{3graphs})~: nearly no signal in $x$ direction ($A_x \approx 0$) and $A_y$ oscillating around $g$ with the double frequency of the pendulum. One can see how at small amplitudes it is 
sinking in the noise while the oscillations can be seeing still very well.

The accelerometer at the rotation axis (central panel in Fig.~\ref{3graphs}) 
shows, as expected, large oscillations from which
the deviation angle can be obtained directly~: $\theta = \arcsin(a_x/g)$. 
The advantage of such measurements particularly becomes evident in the presence of
the pendulum dumping specially introduced to the system to compare large and small
amplitudes in the same figure.

\begin{figure}[!ht]
\centering
\includegraphics[width=6in]{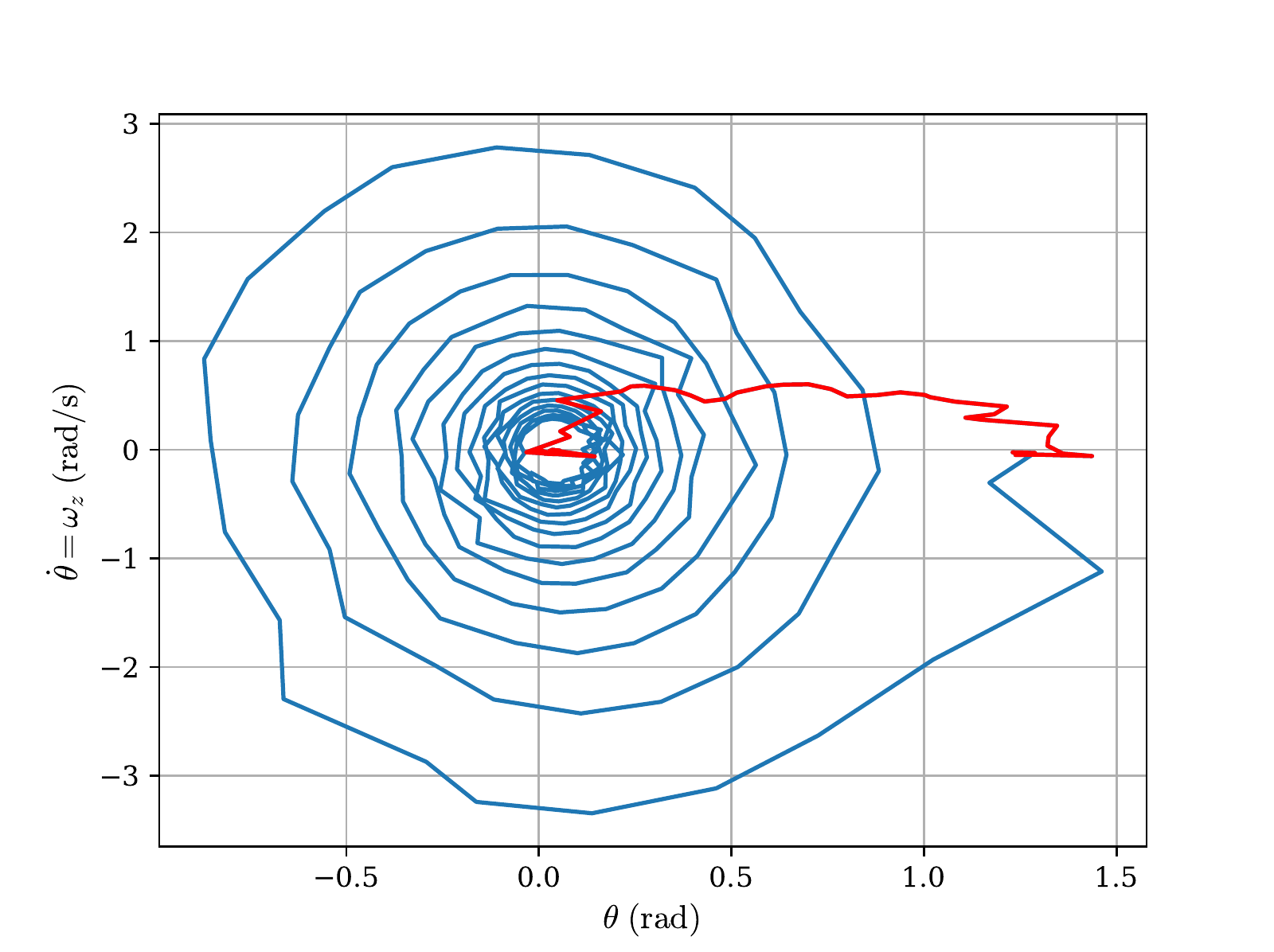}
\caption{Phase portrait of the pendulum oscillation. Red line outlines the initial motion
by hand from the equilibrium position to the release point.}
\label{portrait}
\end{figure}

To complete this experiment we have added the gyroscope of the SensorTag to our program
(lower panel in Fig.~\ref{3graphs})
and obtained $(\theta,\dot{\theta})$ phase diagram (Fig. \ref{portrait}). 
It shows well how, after being put to the initial position (red line in the figure) and released,
the pendulum follows clockwise the spiral of the damped oscillations (blue line). In order
to see this better the damping was increased by adding a helix to the pendulum. Our
$(\theta,\dot{\theta})$ phase diagram is that we find in textbooks unlike
$(\dot{\theta},\ddot{\theta})$ obtained by Monteiro et al.\cite{Monteiro}
One can see how the rotation speed $\dot{\theta}$ ($z$ component measured by the gyroscope 
$\omega_z$) has the maximum when $\theta=0$ and vice verso, all as should be for harmonic oscillations of a simple pendulum.\cite{Boutikov} 

\section*{Conclusion}

In conclusion, we should point out that the proposed measurements with sensors at
the rotation axis give much more clear and direct information on the pendulum motion.
It is also not worse to mention the higher sensitivity for small oscillations
clearly seen.
Moreover, the comparison of the measurements between the two positions makes it possible to highlight surprising counterintuitive phenomena like zero signal observed in the motion
direction ($A_x = 0$) or the double frequency oscillations of the signal $A_y$
along the pendulum.


\end{document}